\documentclass[fleqn,12pt]{SelfArx} 

\definecolor{color1}{RGB}{0,0,185} 
\definecolor{color2}{RGB}{0,20,20} 

\usepackage{hyperref} 
\hypersetup{hidelinks,colorlinks,breaklinks=true,urlcolor=color2,citecolor=color1,linkcolor=color1,bookmarksopen=false,pdftitle={Title},pdfauthor={Author}}
\JournalInfo{Accepted for publication by Astronomy Reports, 2016}

\PaperTitle{
Evidence of the Relationship between the Emerging Magnetic Fields, Electric Currents, and Solar Flares Observed on May~10, 2012
} 

\Authors{M.~A.~Livshits\textsuperscript{1}*,\\ I.~Yu.~Grigoryeva\textsuperscript{2},\\
 I.~I.~Myshyakov\textsuperscript{3}, and\\ G.~V.~Rudenko\textsuperscript{3}} 
\affiliation{\textsuperscript{1}\textit{
Pushkov Institute of Terrestrial Magnetism, Ionosphere and Radio Wave Propagation
of Russian Academy of Sciences, Moscow, Troitsk, Russia
}} 
\affiliation{\textsuperscript{2}\textit{
Main Astronomical (Pulkovo) Observatory, Russian Academy of Sciences, Pulkovo, St.~Petersburg, Russia
}} 
\affiliation{\textsuperscript{3}\textit{
Institute of Solar--Terrestrial Physics, Siberian Branch, Russian Academy of Sciences,
Irkutsk, Russia}} 
\affiliation{* \textbf{Corresponding author}: maliv@mail.ru} 

\newcommand{\keywordname}{Keywords}

\Abstract{We have analyzed multi-wavelength observations and magnetic-field
data for the solar flare of May~10, 2012 (04:18~UT) and have detected a
sign inversion of the signal in the line-of-sight magnetic measurements in
the umbra of a small spot. This effect is associated, at least partly, with
the emergence of a new magnetic field. Almost at the same time, a burst of
hard X-rays was recorded, and 
a ``sunquake'' was generated due to the impact of the disturbance in the
energy release range on the photosphere. At the beginning of the event, a
sigmoid flare was recorded, but it did not spread, as it usually does,
along the polarity inversion (neutral) line. SDO/HMI full-vector
measurements were used to extrapolate the AR 11476 magnetic field to the
corona, and the distribution of vertical currents $j_z$ in the photosphere
was obtained. The distribution of currents in the active region shows that
the relationship between them and the occurrence of flares is very
intricate. We have corroborated that the expected ``ideal'' behavior of the
current system before and after the flare (e.g., see (Sharykin and
Kosovichev, 2015)) is observed only in the sigmoid region. The results
obtained were compared with the observations of two other flares recorded
in this AR on the same day, one of which was similar to the flare under
discussion and the other was of different type. Our results confirm that
the formation and eruption of large-scale magnetic flux ropes in sigmoid
flares are associated with the shear motions in the photosphere and the
emergence of twisted magnetic tubes, as well as with the subsequent
development of the torus instability.
}

\begin{document}

\flushbottom 

{\parindent0pt
\maketitle 
}
\thispagestyle{empty} 
\pagebreak

\tableofcontents 


\section{Introduction}

Over the past 15 years of the new century, TRACE, RHESSI, CORONAS-F, and
other spacecrafts have provided a wealth of data on nonstationary processes
in the Sun. The characteristics of X-ray sources and the relationship with
the microwave and optical emission of solar flares have been studied in
detail (Krucker et al., 2008). In the past 50 years it was believed that
these events occurred in the corona and were due to reconnection of the
magnetic field lines. However, the recent Hinode, SDO/HMI, and AIA
full-vector magnetic observations (Schou, et al., 2012; Lemen, et al.,
2012) and the theoretical extrapolation of photospheric magnetic fields to
the corona (Metcalf, et al., 2008) have shown that, in the course of
evolution of active regions, the magnetic-field energy is accumulated at
very low altitudes in the chromospheres, where the Lorentz forces operate,
and electric currents are amplified significantly. The free energy of the
currents is released at low altitudes giving rise to flares, coronal mass
ejections (CME), and ``sunquakes''. This makes us turn from the recently
introduced term ``coronal flare'' back to the notion of chromospheric
flares, which was widespread in the 1960-1970-ies. This opinion is
supported even by the advocates of the reconnection theory as a mechanism
of flares (Fletcher et al., 2011).

At the early stage of measurements of solar magnetic fields, A.B.Severny
(1988) showed that the sources of individual flare nodes in the
chromosphere are located in the vicinity of the polarity inversion line. He
and his co-authors showed also that these nodes arise in the areas of a
high-magnetic field gradient. The first full-vector measurements of AR
magnetic fields allowed them to study the relationship between the electric
currents and the process of evolution of solar flares. However, the
correlation between the distribution of the vertical magnetic field and the
flare nodes in the chromospheric H-alpha line proved to be very
complicated. Only now the problem discussed by A.B.Severny and D.Rust
(Rust, 1968) can be investigated more or less comprehensively using modern
observations of the magnetic-field dynamics, multi-wavelength observations
of flares, and new theoretical concepts of electromagnetic processes in the
Sun.

Recent studies of the relation between the evolution of magnetic fields and
the nonstationary processes are based on several earlier results. An
important finding was the emergence of new magnetic fields. The change in
the field configuration can result in a sudden release of energy or can
intensify the emission of impulsive flares along with the generation of
plasma motions (CME etc.) and EUV waves (sunquakes). The heating of plasma
in the chromosphere may be due to the dissipation of currents (Zaitsev and
Stepanov, 2008), to the Lorenz forces (Fisher, et al., 2012), or to the
energy loss of accelerated particles at the beginning of the gas-dynamic
response considered first by Kostyuk and Pikelner (1975). To understand the
origin of nonstationary processes, it is necessary to note that in some
cases before or during the flare a large-scale flux rope emerges usually in
the vicinity of the polarity inversion line leading to the occurrence of a
sigmoid flare (see (Hood et al., 2012, and references therein) and to
subsequent formation of post-eruption arcades.

Thus, the appearance of new theoretical and observational data required new
concepts of the development of MHD processes in the outer atmosphere of the
Sun. These concepts should take into account the direct effect of currents
on the heating and motion of plasmas in the chromosphere and its behavior
in the force-free and potential fields in the corona, including the
possibility of reconnection of the field lines and formation of thin
current sheets. Of primary importance in approaching this problem as a
whole is the simulation of the AR current system. In doing so, we must take
into account the topology of magnetic fields in the AR or in the complexes
of activity, which was studied first by Lee et al. (2010). A sudden
reconstruction of the magnetic configuration proves to be closely related
to the development of nonstationary phenomena. The role of shear motions in
such reconstruction was considered earlier (Matyukhin and Tomozov, 1991,
see following discussion in Prist and Forbes, 2000) usually for the
processes with reconnection in the corona. According to the numerical model
by Hood et al. (2012), the emergence of twisted flux ropes is indicative of
such reconstruction of the magnetic configuration and current system.

Of course, it is important to find observational evidence that would
support or disprove the new theoretical conjectures. Perhaps one of the
first publications in this context was the work by Sharykin and Kosovichev
(2015), where Figure 10 represents schematically a current system with
vertical currents $j_z$ of opposite signs on either side of the polarity
inversion line. Indeed, such a picture would be expected in the simplest
case of transition from the twisted flux rope to the flare post-eruption
phase. Now, however, there are numerous full-vector measurements in many
active regions where the distribution of the vertical current $j_z$ is
restored. As in the pioneering work by Severny, this distribution
throughout the AR does not correspond to the ideal pattern.

In this work we made an attempt to investigate the relationship between the
currents and the development of non-stationary processes on the example of
the events on 2012 May 10. The first flare with the maximum at 04:18~UT
(GOES) was unusual since the emergence of the new  field in it was observed
in the umbra of the small spot with the simultaneous hard X-ray burst. This
event was accompanied by a sunquake. The second flare with the maximum at
05:10~UT was more typical, and the third flare (20:26~UT) was like the
first one, but less powerful. After the Introduction, in Section 2 we
analyse the emission of the first flare. In Section 3 the analysis of the
observations of the magnetic field emergence at 04:00~UT is given, in
Section 4 the behavior of the photospheric and coronal magnetic fields is
discussed, and in Section 5 the data are briefly compared with that on the
two subsequent flares. In the last section we discuss the question of the
origin and the possible model of sigmoid flares with the ejection of large
twisted magnetic ropes.

\section{Electromagnetic Emission in the Flare of May~10, 2012 (04:18~UT)}

AR 11476 appeared on the disk on May~5, 2012, and by May~10, its area
reached 1000 m.v.h. The corresponding sunspot group consisted of a large,
complex leading spot and minor spots at the center and in the tail part of
the group. On the day under discussion, the AR produced two large M-flares
and a series of weak flares of class C. Here, we consider the first flare
M5.7 (04:11--4:23, maximum at 4:18~UT, GOES) and then compare it with the
T7.9 flares that occurred at the decay of the first one and with the M1.7
flare observed in the same AR in the evening of that day (see Fig.~1).

\begin{figure}[!t]                                              
\setcaptionmargin{5mm}

\begin{center}
\begin{tabular}{cc}
 & \includegraphics[width=0.3\textwidth,clip=]{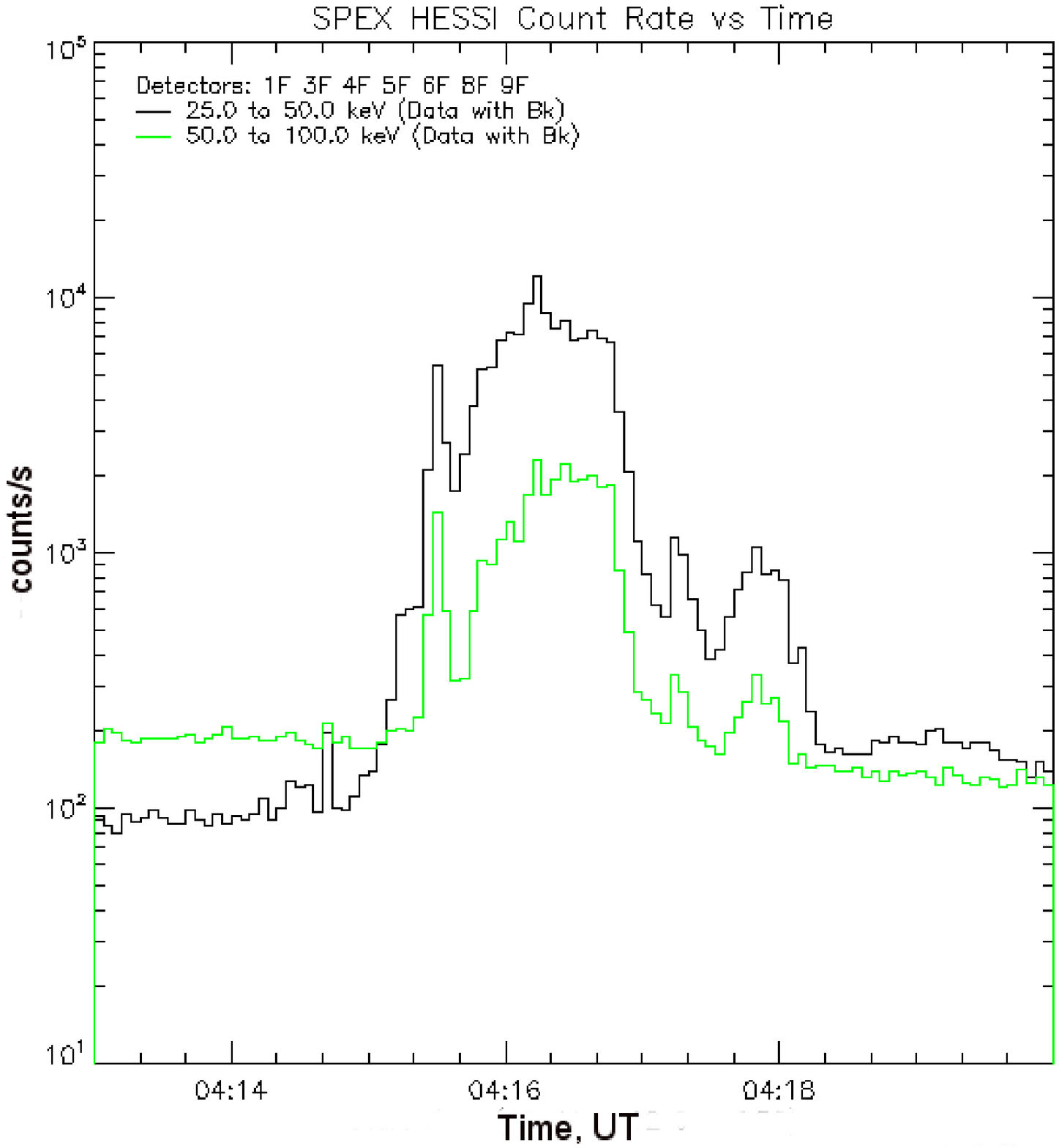}\\
\raisebox{1cm}[0cm][0cm]{\includegraphics[width=0.7\textwidth,clip=]{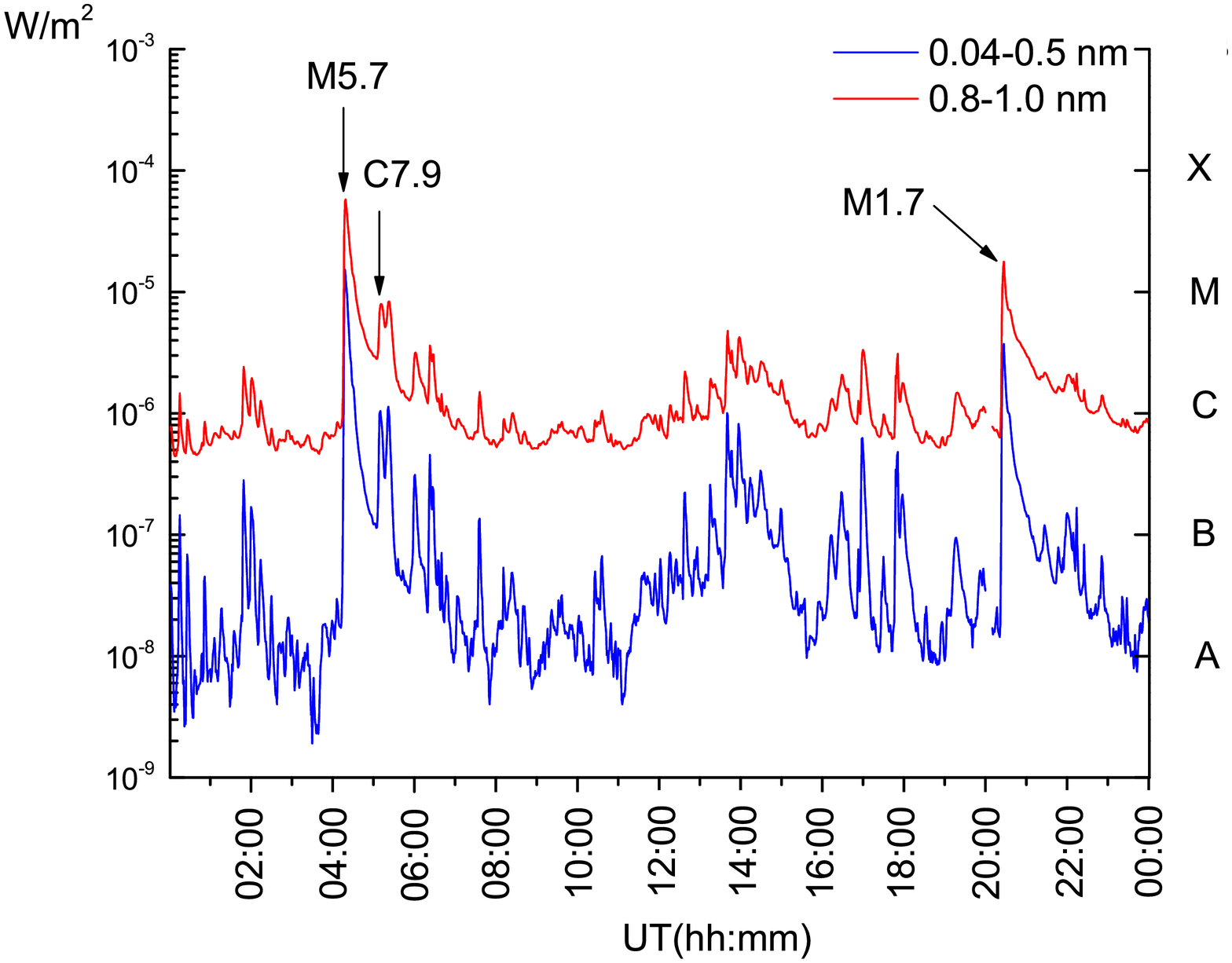}}
 & \includegraphics[width=0.3\textwidth,clip=]{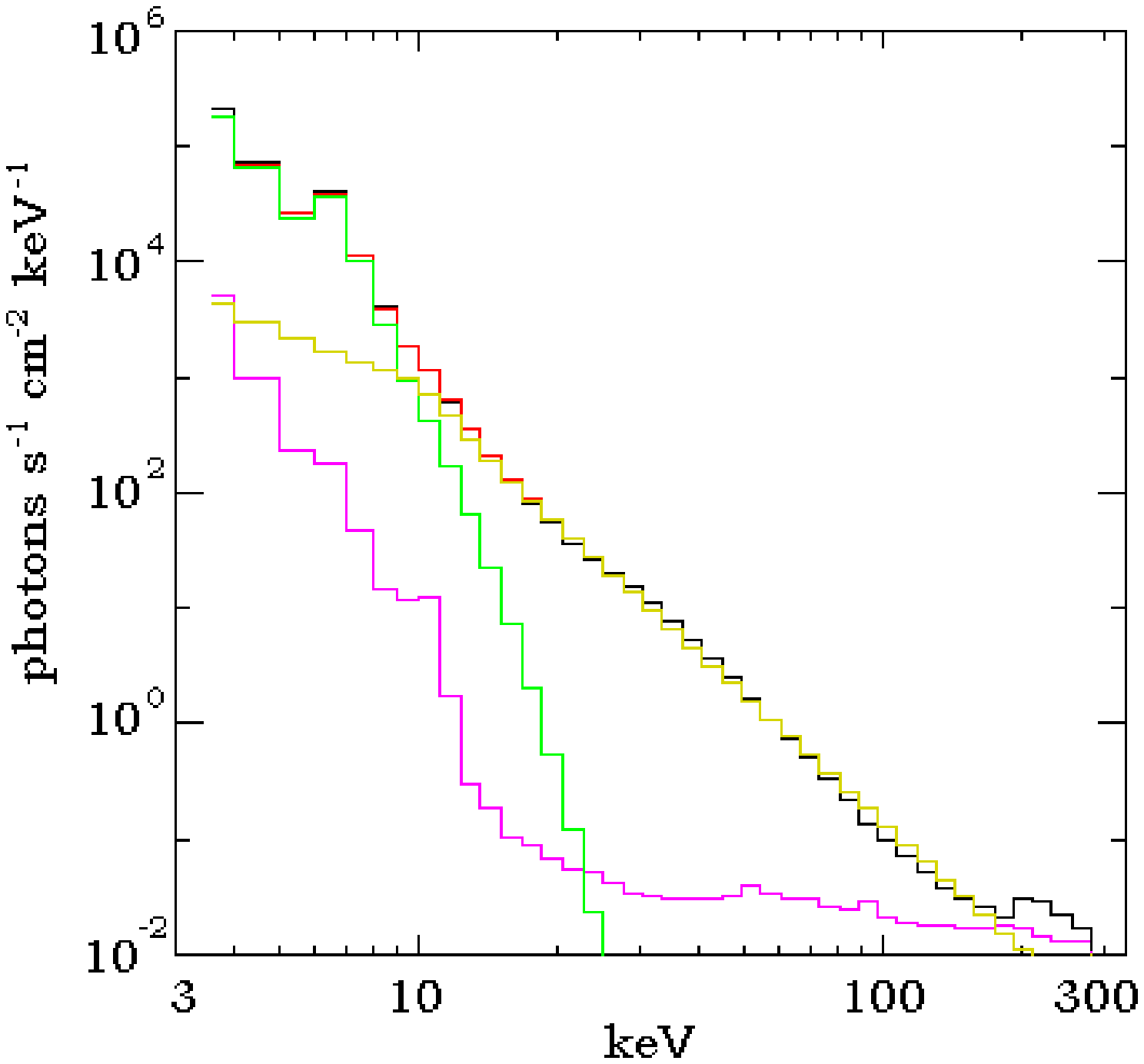} \\
\end{tabular}
\end{center}

\caption{Left: diurnal GOES SXR time profiles (the lines show the flares
under examination). Right, at the top: the RHESSI HXR time profile-- count
rates above the background for the photon energy in the range 25 -- 50~keV
(black line) and 50 -- 100~keV (green line), at the bottom -- the flare
spectrum at 04:15:44~UT on May~10, 2012 where the background is shown as
magenda line. The flare emission is divided into the thermal (green line)
and nonthermal (orange line) components.}\label{fig1}
\end{figure}

The first flare was ordinary as to its X-ray and microwave emission. It
occurred in close proximity to a small spot at the center of the group near
the polarity inversion line, had impulsive nature, and was rather hard.

The left-hand part of Fig.~2 represents the entire AR 11476 in the white
light (SDO/HMI\textunderscore Ic). The rectangle on the right shows region
(A), where the flare was observed, and the squared region (B) is where a
change of sign of the signal was recorded on the line-of-sight magnetogram
(SDO/HMI\textunderscore LOS). The same flare in the 131 \AA\ line (SDO/AIA)
at the growth of its intensity is also shown on the right.

\begin{figure}                                              
\setcaptionmargin{5mm}
\begin{center}
\begin{tabular}{cc}
&\includegraphics[width=0.45\textwidth,clip=]{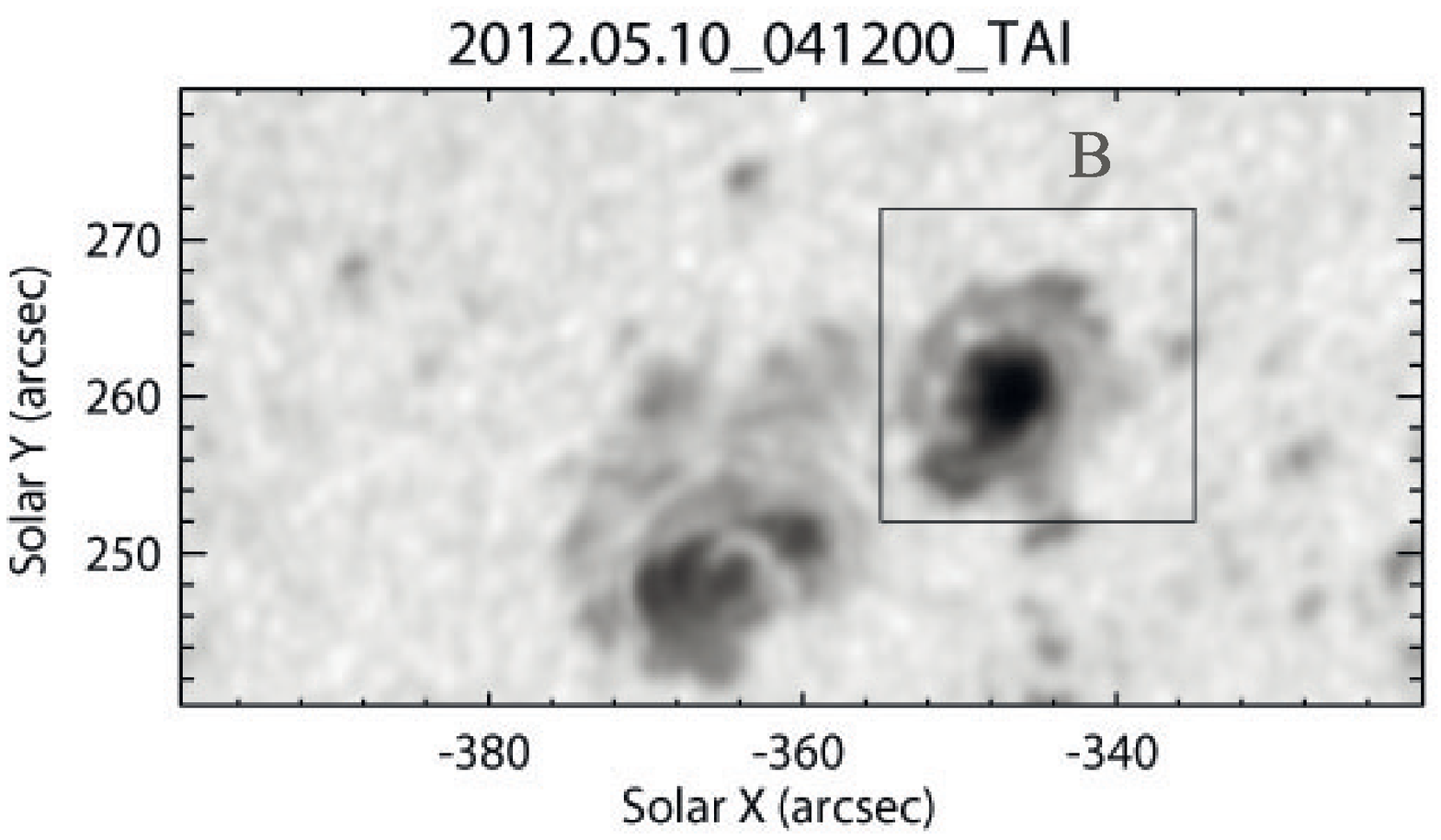}\\
\raisebox{0.1cm}[0cm][0cm]{\includegraphics[width=0.55\textwidth,clip=]{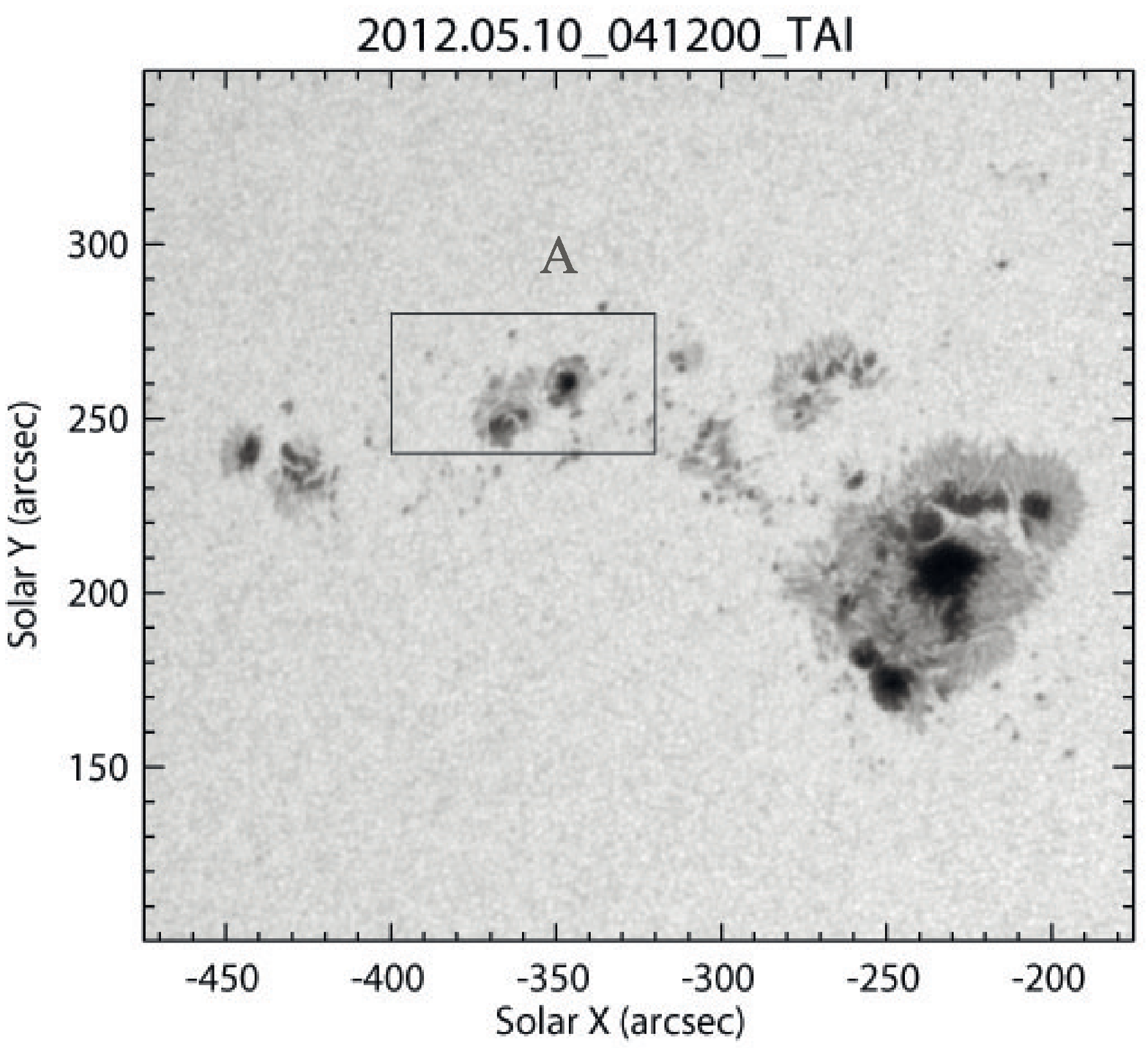}}&
\includegraphics[width=0.45\textwidth,clip=]{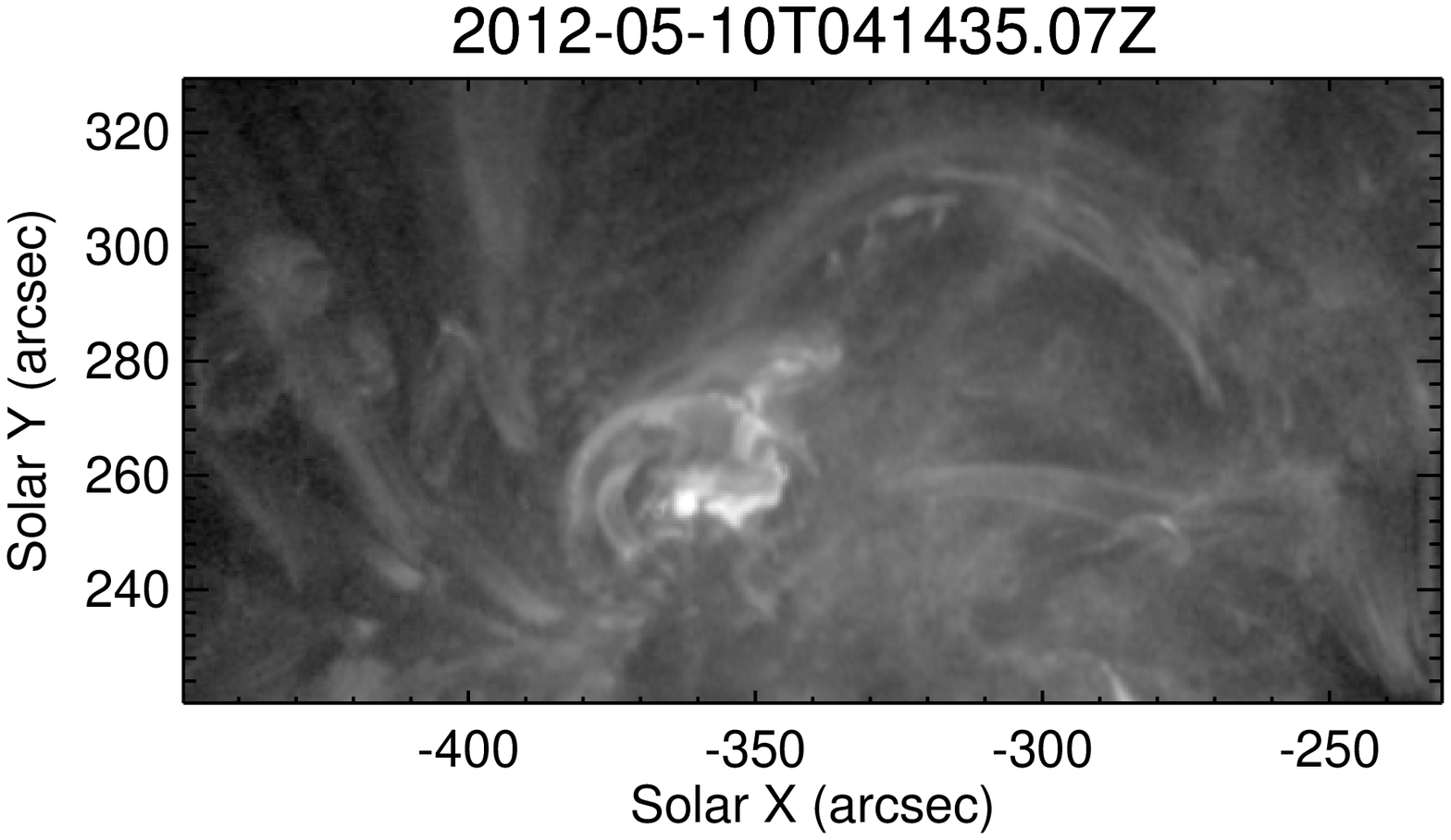}\\
\end{tabular}
\end{center}
\caption{Left: the sunspot group in AR 11476 at 04:12~UT
(SDO/HMI\textunderscore Ic), the rectangle is the selected region A. Right:
region A with the square region B for the same time (at the top); flare in
the 131 \AA\ line (SDO/AIA) in the vicinity of the maximum (at the bottom).
The scales show the distance from the disk center in arcsec.}\label{fig2}
\end{figure}

In general, judging by its X-ray emission recorded by GOES and RHESSI (Lin,
et al., 2002), this was a typical event somewhat harder than an average
flare of importance M5 as seen from Fig.~1 (the tilt of the photon spectrum
$\gamma\approx 3.0$ in the energy range of 20--80~keV, Fig.~1).

The microwave flux was also characteristic of the flares of this class with
a complex behavior of the polarized component (Stokes parameter V). The
time profiles of intensity (Stokes parameter I) and polarization (Stokes
parameter V) at the frequencies of 17.0, 9.4, 3.75 and 2.0~GHz based on the
Nobeyama polarimeter data (NoRP: Shibasaki et al., 1979;
Nakajima et al., 1985) are given on the left and at the center of Fig.~3.
The values and the dynamics of fluxes correspond to the majority of flares
of this intensity. As usual, there are two discernible footpoints of the
flare loop, located in the magnetic fields of the opposite signs (Fig.~3,
right).

\begin{figure}                                              
\setcaptionmargin{5mm}
\begin{center}
\begin{tabular}{ccc}
\includegraphics[width=0.3\textwidth,clip=]{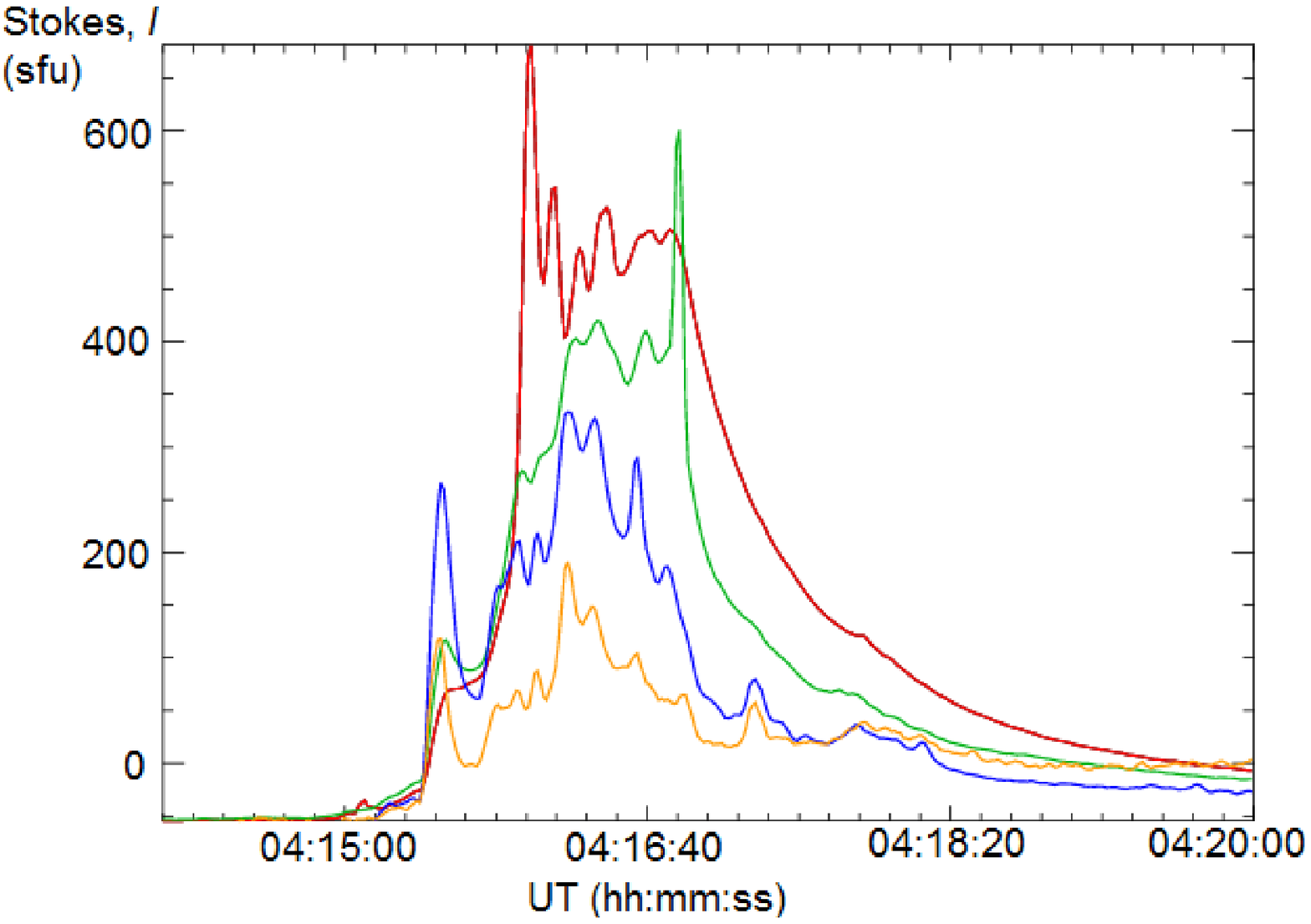} &
\includegraphics[width=0.3\textwidth,clip=]{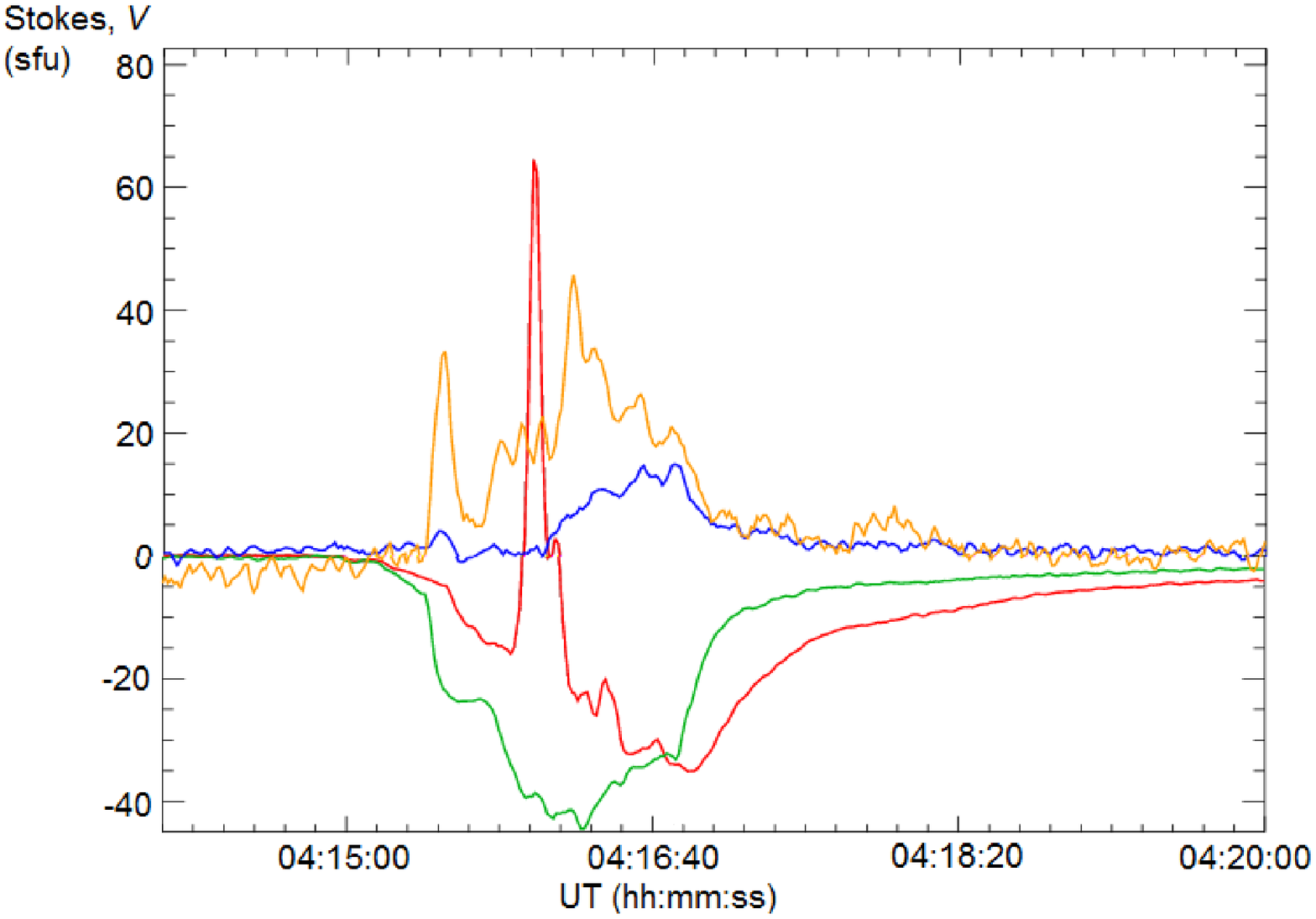} &
\includegraphics[width=0.3\textwidth,clip=]{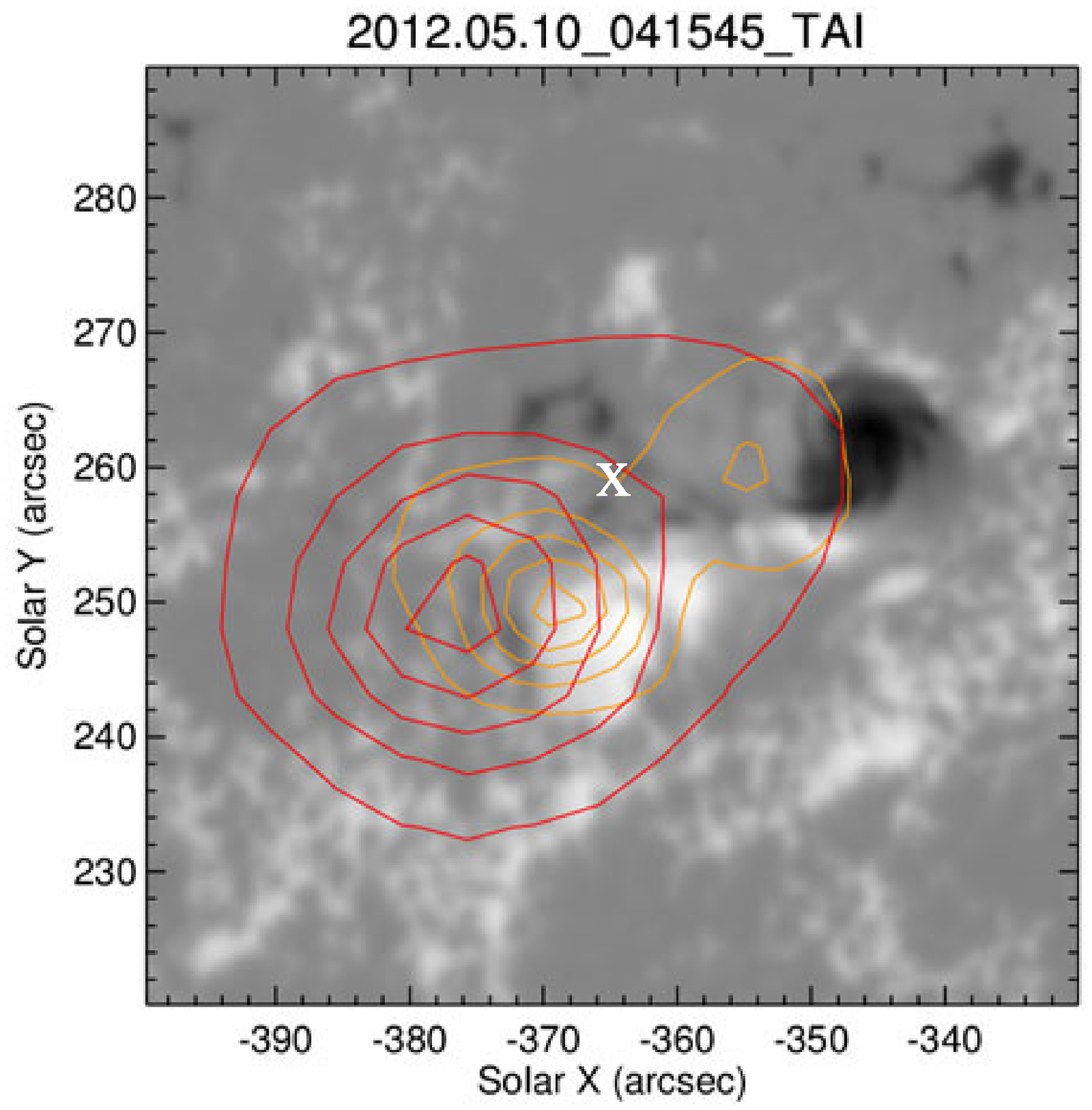}
\end{tabular}
\end{center}
\caption{Evolution of the microwave flux from the M5.7 flare on May~10, 2012
(Stokes parameters I (left) and V (in the centre) at the frequencies of 2
GHz (red) 3.75~GHz (green), 9.4~GHz (blue) and 17~GHz (yellow) (Nobeyama,
NoRP).
Right -- a fragment of the LOS-magnetogram of the flare generation region
in the phase of maximum: the scales provide the distance from the disk
center in arcsec; the contours correspond to the emission at 34.0 GHz
(yellow line) and 17.0 GHz (red line) (Nobeyama, NoRH) at 04:15:30~UT;
these isophotes correspond 90\%, 70\%, 50\%, 30\% and 10\% of the
maximum at 17 GHz and 34 GHz. The sign X shows the flare occurrence site
according to HXR RHESSI data.}\label{fig3}
\end{figure}

On the other hand, the polarization demonstrates a rather complicated time
behavior.  In particular, In particular, we have to be an inversion of the
polarized emission sign at the frequency of 9.4 GHz. Besides, a sudden
change of sign is observed at 2~GHz at 04:16~UT. Such a picture of
evolution of the polarized microwave emission may  be associated with a
complex AR topology in the lower corona.

\section{Emergence of Magnetic Fields}

It is well known that in many cases, a relationship can be observed between
the occurrence of a flare and the emergence of a new magnetic field.
However these two processes do not usually coincide in time and space. In
other words, the emergence of a new magnetic field violates the stability
of the magnetic configuration. The flare in a given AR may occur far enough
from the emerging field, before or after the emergence or, even between its
separate episodes. In the case under consideration, the two processes
coincided in space and time. Besides that, though usually the new magnetic
field appears in the vicinity of the polarity inversion line separating the
fields of medium-high intensity, in our case, the emergence occurred in the
umbra of a small spot at the AR center (square B in Fig.~2). Here, the
change of sign of the signal (SDO/HMI\textunderscore LOS) is indicative of
the field emergence, but still it certainly does not mean the inversion of
the field sign.

We have analyzed the Stokes parameters recorded in the supposed emergence
area and obtained the Gaussian distribution. Of course, strong plasma
streams may also interfere with the field measurements. One of the
arguments for the reality of the field emergence is the fact that a similar
picture was observed during the flare recorded at the same site in the same
AR later in the evening at 20:26~UT.

The evolution of the emerging field in region B at some selected points of
time based on the \sloppy SDO/HMI\textunderscore LOS magnetograms is illustrated in
Fig.~4. The reversed sign of the signal was observed during 4 min., after
which the signal was restored to its former value during another 4 min.
Unfortunately, full-vector magnetic data are only available for the moments
before and after the emergence. Note that the upward directed motions are
registered according to the data of SDO/HMI in the frame 04:16:30~UT.

\begin{figure}                                              
\setcaptionmargin{5mm}
\includegraphics[width=0.9\textwidth,clip=]{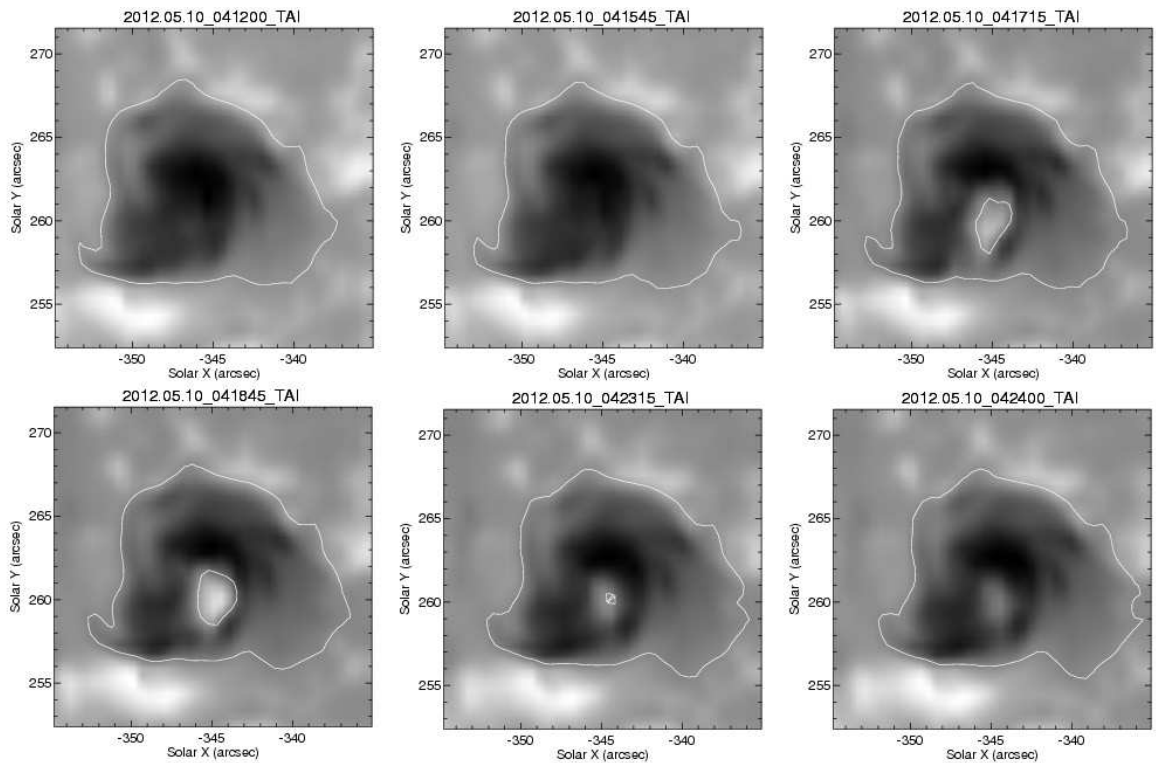}
\caption{Fragments of the image of the magnetic-field line-of-sight
component in the emergence region in AR NOAA 11476 (SDO/HMI\textunderscore
LOS) on May~10, 2012 at the selected moments during the flare.   The
contours show the isolated local magnetic field along the polarity
inversion line. The scales provide the distance from the center of the disk
in arcsec.}\label{fig4}
\end{figure}

During the event under discussion, a sunquake was recorded in the vicinity
of the hard X-ray maximum (Buitrago-Cass, et al., 2015). Such a response to
the sudden energy release in the photosphere is projected onto the point at
the peak of a low flare loop. The asterisk (X) in the right-hand part of
Fig.~3 marks the position of the hard X-ray source at the maximum of the
burst. The isolines of radio brightness at 17.0 and 34.0 GHz superimposed
on the line-of-sight distribution of magnetic fields at the maximum of the
event are given according to the Nobeyama radio heliograph data (NoRH:
Nakajima et al., 1994). The footpoints of the low flare loop seen in
different SDO/AIA ranges are best pronounced at the high frequency of 34.0
GHZ. In this case, the sunquake is most likely bound to the time and site
of the primary energy release. This suggestion agrees with the sunquake
data available in (Sharykin and Kosovichev, 2015; Sharykin, Kosovichev, and
Zimovets, 2015)

\section{Extrapolation of Fields to the Corona and Distribution of Electric Currents in AR 11476}

In recent years, our understanding of the role of magnetic fields in the
development of activity in the outer atmosphere of the Sun has changed. It
was found that the magnetic beta (the ratio of the gas pressure to the
magnetic pressure), which is close to unity immediately above the
photosphere, becomes very small in the chromosphere and corona. As a
result, the plasma at the base of the active region is exposed to the
Lorenz forces, while higher the field becomes force-free. At the same time,
the evolution of fields in the chromosphere can lead to a significant
excess of the magnetic-field energy over the corresponding value of the
force-free (potential) field, and this free energy can be spent on the
development of nonstationary processes.

Full-vector magnetic observations in the photosphere allow us to
extrapolate the nonlinear force-free fields up to the corona. This method
called NLFFF (nonlinear force-free field) extrapolation is widely used in
heliophysics (e.g., see (Metcalf et al, 2008)).

In our work, for the calculation of the coronal magnetic field in the
non-linear force-free approximation an optimisation method (first proposed
in (Wheatland et al, 2000)) is employed in the implementation of
(Rudenko and Myshyakov, 2009). The core of the method is the subsequent transformation
of some initial (potential) distribution of the magnetic field toward the
force-free structure in accordance with the photospheric magnetogram. To do
this a functional of the following type is introduced:

We have performed calculations based on the solution of the
boundary-(Rudenko and Myshyakov, 2009) value problem for the nonlinear
force-free field using minimization of the functional introduced in:
\begin{eqnarray}
L=\int_V[B^{-2}\vert[\nabla\times\mathbf{B}]\times\mathbf{B}\vert^2+
\vert\nabla\cdot\mathbf{B}\vert^2\,] dV,
\end{eqnarray}
which equals $0$ if the field $B$ is force-free and is positive in the opposite case.

Thus, in the course of the solution of the functional (1) minimisation
problem, the field in the volume $V$ acquires the force-free configuration.
As the initial distribution we use a potential field calculated from the
normal component of the photospheric field via fast Fourier transform
(FFT). A remarkable feature of the optimisation method's realisation, which
we apply, is the use of the full system of the evolution equations of the
field (see (Rudenko and Myshyakov, 2009)).

The field lines in region A shown in Fig.~2 were calculated up to the
height of 25\,000~km. The result is represented in Fig.~5. One can see that
at the very beginning of the flare, the force lines around the field hill
in a small spot are clustering along the polarity inversion line, and the
flare nodes are formed in the region of their highest concentration (see
the EUV background on the negative image in Fig.~5). Generally speaking, a
bunch of force lines covering a significant part of the polarity inversion
line, particularly, where it separates the hills of strong magnetic fields,
can form in many flares. In our case, it can be supposed that a small bunch
was formed only on the south side of a small spot at the center of the
group. As in other sigmoid flares, such a shape of the field lines in AR
stimulates the development of sigmoid flare. In this case, the field
reconstruction occurred almost immediately after the formation of currents
over the neutral line.

\begin{figure}                                              
\setcaptionmargin{5mm}
\includegraphics[width=0.95\textwidth,clip=]{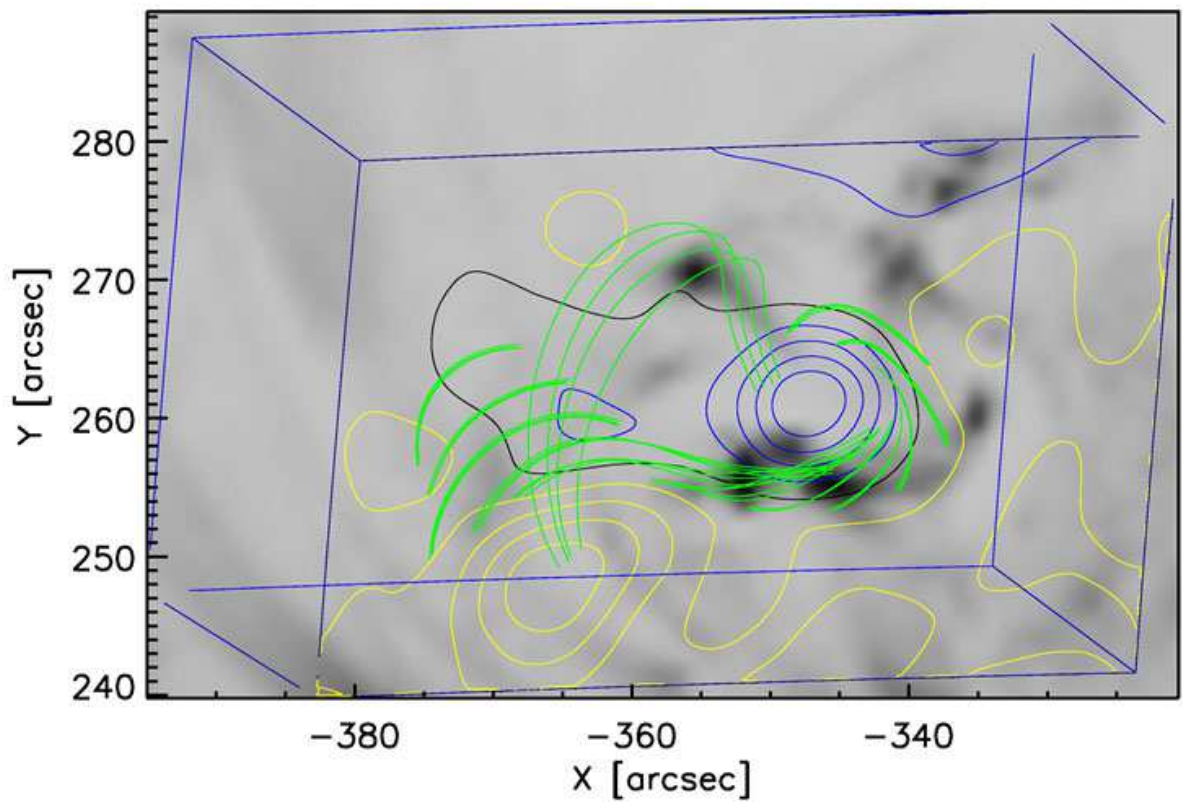}
\caption{The background is a fragment of the inverted image of AR NOAA 11476
(SDO/AIA) in the EUV line 171 \AA\ obtained on May~10, 2012 at 04:12:00~UT.
The contour lines show the line-of-sight components of the magnetic field
of positive (white) and negative (black) sign in the flare region at 80\%,
60\%, 40\%, and 20\% of the maximum at the same time. The white line
presents force lines above the neutral line and the higher loops connecting
hills of the magnetic field of the opposite signs. The direct blue lines
show the projection cube with the height of  25\,000~km. The scales provide
the distance from the center of the disk in arcsec.}\label{fig5}
\end{figure}

Figure 5 shows only closed field lines in the selected region A in
projection on the plane of the sky. We can see a small spot of negative
polarity fringed with a bunch of field lines that continue further beyond
the neutral line. At present it is believed that this is where the main
force-free currents usually identified with the plasma flux rope are
flowing. It is important to note that this bunch of field lines coincides
with the position of the flare nodes forming a small sigmoid.

As shown by NLFFF-extrapolation, there are also field lines rising up to
5000 km and higher and connecting the small spot of negative polarity with
the hill of the field of opposite sign. They are located above the bunch of
field lines and perhaps prevent the ejection. In the course of the flare
evolution as inferred from SDO/AIA EUV data, the flare loops 10-20 thousand
km high were formed, but plasma did not escape into interplanetary space
(i.e., a CME did not occur as it follows from the radio data: there is no a
type II burst etc.).

Then at the photosphere level we have calculated the distribution of
vertical currents $j_z$ throughout the AR on May~10, 2012 for the time
interval from 03:48~UT to 05:48~UT using the full-vector magnetic data with
a step of 720 s. The analysis of the location of flares on that day
corroborates the correlation between the sites of occurrence of the flare
nodes and the areas of enhanced current (Grigor'eva et al., 2013). However,
no definite regularity was revealed in the behavior of the currents
throughout the AR and in the occurrence of flares. Namely, all currents are
concentrated along the neutral line, including the area near spots.
Therefore, the changes that may occur in the middle of the neutral line in
AR are not identified in the total current of one and the other sign.
Sharykin and Kosovichev (2015) drew attention to the fact that in an
``ideal'' current system, the mean currents of each sign must increase on
either side of the polarity inversion line. In our case, like in the
studies of the 1970-ies considering the relationship between the vertical
currents and flares, this effect is very weak in respect to the entire AR.

Since in the growth phase of the flare the largest changes in the magnetic
field are recorded at the site of occurrence of the sigmoid, we have
studied this effect in the selected region A (see Fig.~1). Figure 6
illustrates the distribution of the vertical currents in the photosphere
for two points of time. One can see that the currents of different sign
contact at the point  $Y~256$, $X~-350$  arcsec. This area coincides with the
region of the field lines concentration in Fig.~5 above the point on the
polarity inversion line in the region of strong fields (see
Fig.~\ref{fig4}).

Generally, the evolution of the currents pattern is traced upon the set of
the $j_z$ maps constructed on the basis of the data on the full vector of
the field. The largest change of the currents takes place in the pointed
region of the sigmoid localization. The subsequent evolution of the current
system appears to be tightly bound with the development of numerous weak
flares in AR~11476. Note also that the substantial change of the currents
took place at  04:00--04:24~UT in the area $Y~250$, $X~-364$ arcsec which is
the footpoint of the loop connecting the place of the magnetic field
emergence with the hill of the positive polarity (see Fig.~\ref{fig5}).

\begin{figure}                                              
\setcaptionmargin{5mm}
\begin{center}
\begin{tabular}{cc}
{\includegraphics[width=0.45\textwidth,clip=]{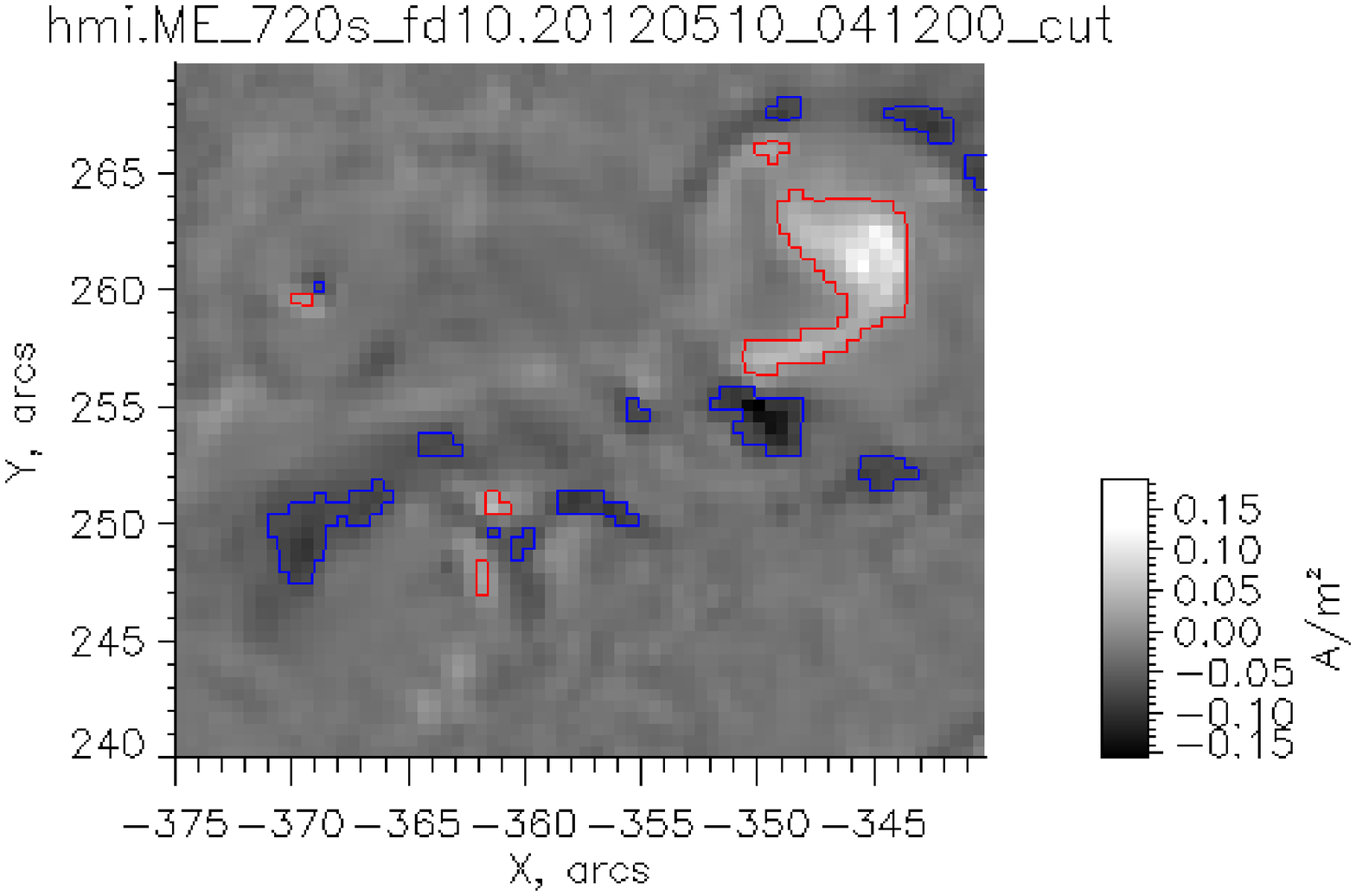}}&\\
{\includegraphics[width=0.45\textwidth,clip=]{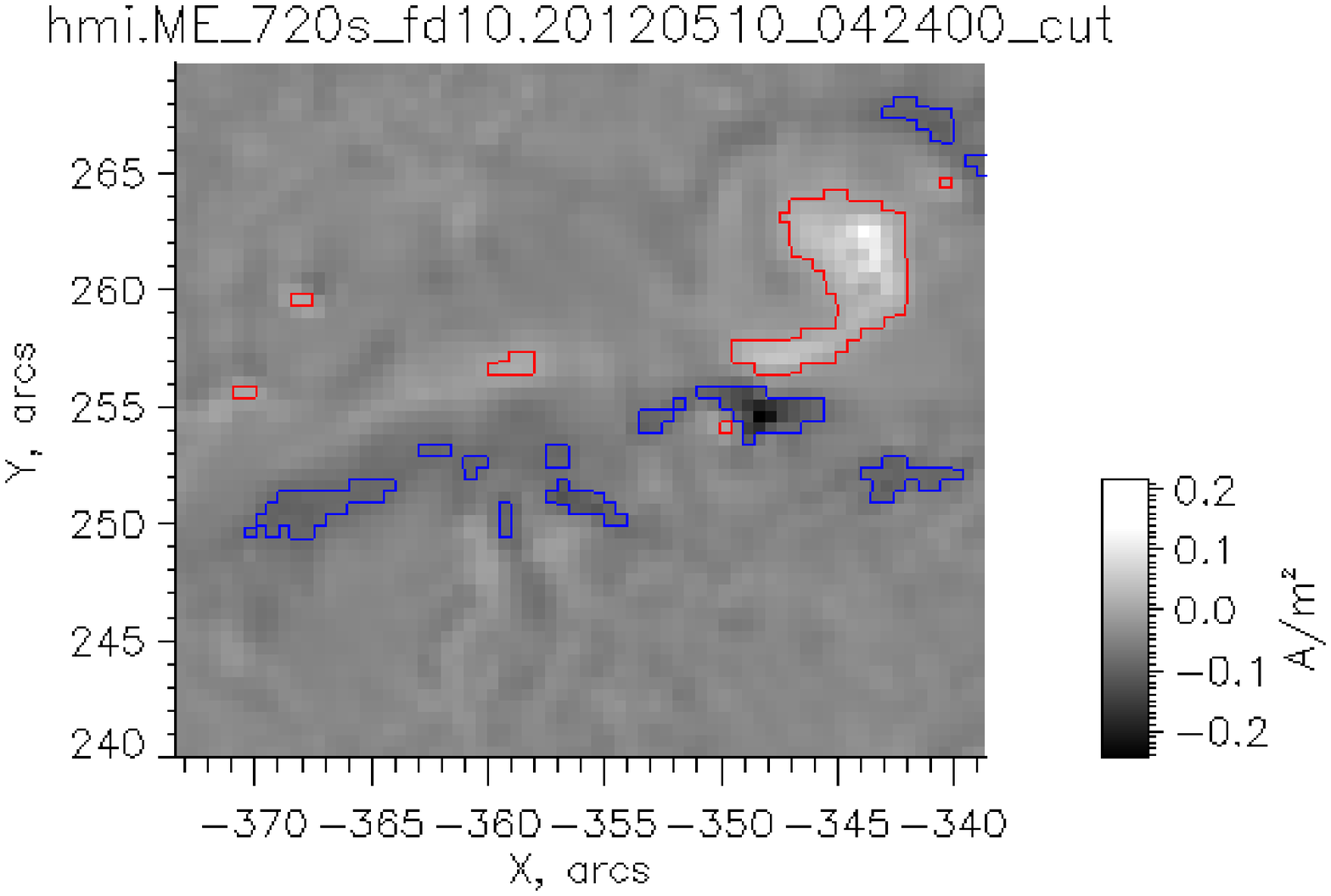}}&
\raisebox{2cm}[0cm][0cm]{\includegraphics[width=0.55\textwidth,clip=]{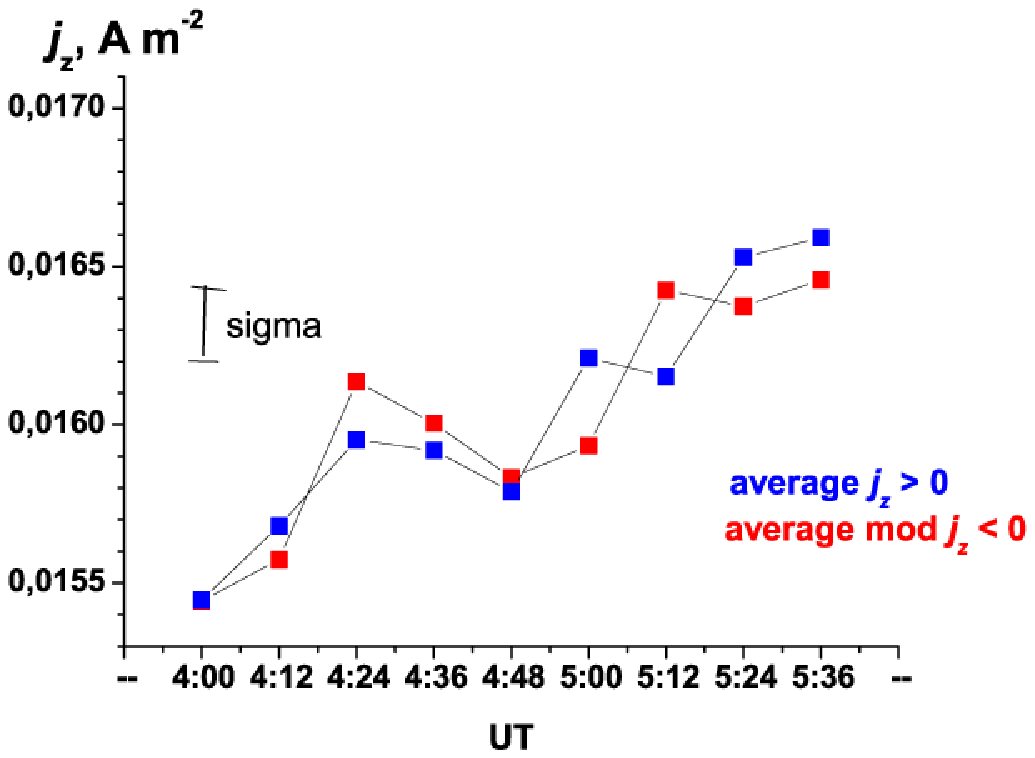}}\\
\end{tabular}
\end{center}
\caption{Left: (the background) fragments of the maps of vertical currents,
$j_z$, calculated for 04:12~UT and 04:24~UT, i.e., before and after the
flare maximum, respectively; (the contours) levels $j_z\pm0.05$~A/m$^2$ for
$j_z<0$ (red color) and $j_z>0$ (blue color). The scales show the distance
from the disk center in arcsec. Right: evolution of the mean current
$j_z>0$ and $j_z<0$ on May~10, 2012. For the magnetic field of negative
sign, the modulus of~$j_z$ is shown. The scale provides the time in the
form h:mm~(UT).}\label{fig6}
\end{figure}

Except the change of the currents pattern, the difference between two
images on left part of Fig.~\ref{fig6} is manifested in all the current
values, the fact which in this example is revealed even in the difference
of the scales.

To study this effect, we have determined the mean value of the vertical
current in the whole region (A). The averaging was done separately for the
pixels with the positive and negative fields. In the time interval from
04:00~UT to  05:48~UT in Fig.~\ref{fig6} one can observe the general trend
of the averaged vertical current $j_z$ of both signs. The whole trend
represents the fact that the flaring activity of this AR on that day was
gradually increasing.

The effect related to the flare in question is manifested as the increase
of the averaged currents during the flare  04:00~UT--04:24~UT. According to
the hard X-ray data, the maximum of the flare falls on 04:16:30~UT (see
Fig.~\ref{fig1}). In our case the maximum value corresponds to the time
04:24~UT and falls down afterwards. Such character of the change of $j_z$
is noticed in (Sharykin and Kosovichev, 2015) as well. The maximum value is small and
only slightly exceeds the value of 3~$\sigma$ (its  $\sigma$ value of
approximately $2.3 \cdot 10^{-4}$~A/m$^2$ is determined from the change of the
currents in the area of the same active region in the quiescent state).

Note also that the flare maximum pointed out occured between 04:12~UT and
04:24~UT. Unfortunately, there is no freely accessed data on the full
magnetic field vector with better time resolution, therefore our effect of
the change of the currents' characteristics during this flare can be
somewhat underestimated.

\section{The Other Flares of May~10, 2012}

AR 11476, which we are discussing, displayed a high flare activity. Every
day since May 6 it produced a large number of weak C-flares and, on some
days until May 17, 1 or 2 M-flares.

In addition to the flare of importance M5.7 we are considering here, it
produced on the same day 16 C-flares and one flare of importance M1.7
recorded at 20:20~UT. A change in the signal intensity was recorded in the
umbra of the same spot as in the first M-flare according to the
line-of-sight field distribution (SDO/HMI\textunderscore LOS). The duration
of this effect was two times less than in the first flare (from ~20:24~UT
to 20:29~UT). However here, unlike the M5.7 flare, the signal on the
HMI\textunderscore LOS magnetogram in the emergence region did not change
the sign.

This M1.7 flare was also rather hard. The FERMI space mission recorded the
emission in the range of about 100 keV. At the same time (about 20:24~UT),
significant radiation fluxes in the energy range above 50 keV with a
maximum of more than 2000 pulses per second were recorded by the Suzaku
spacecraft (see the catalogue 
\href{http://www.astro.isas.jaxa.jp/suzaku/HXD-WAM/WAM-GRB/solar/untrig/120510202227.html}{Suzaku Wide-band All-sky Monitor (WAM)}:
{\footnotesize{\underline{ http://www.astro.isas.jaxa.jp/suzaku/HXD-WAM/WAM-GRB/solar/untrig/120510202227.html }}}).
Thus, the conditions that lead to the events of the type of a ``fast''
sigmoid can arise in AR with a period from several hours to a day.

Most of the flares of class C on the day under discussion had a different
origin than the M-flares. Let us consider by way of example the flare T7.9,
which was observed at the decay of the soft X-ray emission from our
principal event. The flare occurred at 05:04~UT over the polarity inversion
line between the spots we are considering and the leading spots. Then, the
flare nodes extended from this point directly to the large spot. Such weak
events might be triggered by the activity in the latter. In some cases,
this activity is due to the emergence of a new field in the vicinity of the
spot or with its rotation, or simply with the formation of new magnetic
hills. Some of the weak events are rather hard, but they occur usually at
the decay of the soft X-ray emission from major flares. Their hardness is
due to the conditions favorable for particle acceleration in the traps
where a certain number of electrons with energies above 10 keV still remain
after the previous event (Vybornov et al., 2015).

\section{Results and Discussion}

Our conclusions are as follows:

1) In the event under examination, there actually occurred a pulse-like
energy release accompanied by the bursts of hard X-ray and microwave
emission. Simultaneously, a response of the photosphere was recorded in the
form of an sunquake. There is every reason to believe that at
the same time, the formation of a sigmoid flare began, but it was not
completed. Similar processes were repeated in the second M-flare at
20:20~UT, but were absent in many weak flares of class C in the other parts
of this AR. 

2) Full-vector magnetic data were used to calculate the
vertical currents throughout the AR. A simple pattern of the current system
that consists of the loop currents above the neutral line (with allowance
for the shear) and the currents closed under the photosphere does not agree
with the results of calculations for first flare. However, if such
calculations are carried out in an area confined to the size of the flare
sigmoid, we will find local maxima in the time variation of both positive
and negative vertical currents. For the longer time interval, the time
variations of averaged vertical currents determines the overall evolution
of the flare activity in a given AR. 

3) This example of moderate power
event has demonstrated at both the sunquake and particle acceleration are
more effective in sufficiently strong magnetic fields residing near the
polarity inversion line.

Let us briefly discuss the connection of our results to the gentral problem
of the development of non-stationary processes. The phenomenon of sigmoid
flares has been widely discussed in literature. A rich experience of
numerical simulation of these processes has been accumulated (e.g., see
(Hood et al., 2012, and references therein). Formerly, many authors adhered
to the idea of reconnection of magnetic-field lines. Now, it is believed
that the main factor is the dynamics of the currents, while the role of the
reconnection becomes significant at high altitudes in the corona. The
earlier conclusion that plasma flux ropes with the current are not ejected
directly from the top of the convection zone remains valid. Large-scale
sub-photospheric motions shear the feet of the field lines along the
polarity inversion line. In addition to that, the helicity is also carried
out from sub-photospheric layers.

So, the emergence of the shear and helicity creates conditions for the
formation of a magnetic flux rope. Both theoretical reasoning and
extrapolation of the nonlinear force-free fields to the corona suggest that
the evolution of the AR magnetic field described above must result in
accumulation of some amount of free energy in the chromosphere, which can
be spent on the development of nonstationary processes.

The very moment of a large impulsive flare is associated with the
reconstruction of the current system in AR. Namely, the bunch of force
lines and the currents along them change essentially approaching the ideal
configuration, i.e., currents along the loops, which are located in
projection onto the photosphere at a large angle to the neutral line.

After the flux rope is released, the magnetic configuration restores its
original state. If the previously existing large-scale sub-photospheric
motions continue, the formation of the flux rope is repeated, i.e., a
series of similar flares occur. These considerations agree with the results
of the latest numerical calculations (Savcheva, 2016).

One important issue is to figure out a mechanism of  the flare. We note
that prior a flare the strong current in the magnetic rope has to be closed
under the photosphere. This suggests the toroid in the meridian plane,
partially risen above the photosphere.  This geometry of  the twisted
magnetic fiekd is given in Figure 2 in Titov and D\'emoulin (1999).

In fact, the flux ropes can exist for a long time even in active regions of
rather complex topology. However, in some cases, especially at geat
currents,  a torus instability can arise, particularly in the presence of
strong currents (Shafranov, 1966). Conditions for the development of such
instabilities were determined in laboratory experiments (Myer et al.,
2015). Of course, for flares on the Sun and low-mass stars, we cannot
insist upon  a specific type of instability, and more resonable to consider
a more general case of hydrodynamical instabilities in the plasma rope with
the strong current.

Simulation of the current system throughout the AR is an important problem.
Only now it has become clear that strong currents and large Lorenz forces
exist very low (up to 2-3 thousand km above the photosphere). Higher in the
corona, the fields rapidly become potential, and thin current sheets are
possibly present.

Another evidence in addition to the general considerations, is a complex
behavior of the polarization of microwave radiation. It is known that the
polarization changes its sign when passing from the disk center to the
limb. This is because the beam enters quasi-transverse magnetic field
(QT-region) (e.g., see Zheleznyakov, 1963; Peterova, 1973). In our case
(see Fig.~3) we see the inversion of the polarization sign at the frequency
of 9.4 GHz at the same time (without motion of the source).

In addition to this well-known effect, a secondary short-time inversion was
recorded at the frequency of 2 GHz with the maximum polarized emission at
04:16~UT. This, in turn, suggests a complex topology of the coronal
magnetic field, when a complex polarization picture is due both to the
projection effects, and to increasing complexity of the coronal topology.

Thus, an ``ideal'' current system can exist within a small volume. But
usually it is immersed into various independent magnetic fluxes separated
by separatrix surfaces. As a result, the particularities of the polarized
emission that appear at low frequencies in the microwave range indicate the
role of reconnection in the coronal layers in AR. Note that the foundations
for investigation of the topology of the solar corona were laid by Lee et
al. (2010).

Note also that two scenarios of events are possible in the Sun: (1) with
coronal mass ejection (CME), frequently observed, particularly, in major
flares and (2) with a rising loop without ejection (rarer events). The
latter was observed in the M-flares on May~10, 2012.

\phantomsection
\section*{Acknowledgments}

We are grateful to the reviewers and to V.I.~Vybornov for his assistance in
the course of the work. The SDO data are the courtesy of NASA and the HMI
and AIA science teams. We acknowledge the use of RHESSI, GOES and Fermi/GBM
data. We are grateful to the instrumental teams operating the Nobeyama
solar facilities and GOES satellites.

The work was supported by the Russian Foundation for Basic Research,
project nos.~14-02-00922, 16-32-00315, 15-32-20504, 15-02-03835,
15-02-01089, 15-02-01077, 16-02-00749, and particularly by Program no.~7 of
the Presidium of the Russian Academy of Sciences.

\phantomsection
\bibliographystyle{unsrt}

\end{document}